\documentclass[twocolumn,pre,tightenlines,nofootinbib,amsmath,amssymb,amsfonts,superscriptaddress,eqsecnum]{revtex4}
\usepackage{subeqnarray}
\usepackage{hyperref}
\usepackage{lipsum}
\usepackage{accents}
\usepackage{yhmath} 
\usepackage{array}
\usepackage{graphicx}
\usepackage{bm}

\newcommand{\be}{\begin{equation}}
\newcommand{\ee}{\end{equation}}
\newcommand{\bea}{\begin{eqnarray}}
\newcommand{\eea}{\end{eqnarray}}

\newcommand{\beas}{\begin{subeqnarray}}
\newcommand{\eeas}{\end{subeqnarray}}

\newcommand{\dd}{{\rm d}}

\newcommand{\hrbp}{{\cal E}}
\newcommand{\hrb}{{\cal R}}
\newcommand{\gr}[1]{{\bm #1}}

\newcommand{\com}[1]{}

\newcommand{\ii}{{\rm i}}
\newcommand{\tetr}[1]{{\underline{#1}}}

\newcommand{\totV}{{\cal V}}

\newcommand{\ra}{{R}}

\newcommand{\visc}{{\mu}} 
\newcommand{\ST}{{\nu}}

\newcommand{\ScV}{\overline{\cal V}}
\newcommand{\Sv}{v}
\newcommand{\Sw}{{\dot{\phi}}}

\newcommand{\tkappa}{{\widetilde{\kappa}}}

\include{RevtexAuthorYear}
\begin{document}
\title{One-dimensional reduction of viscous jets. II. Applications}

\author{Cyril Pitrou}
\email{pitrou@iap.fr}
\affiliation{Institut d'Astrophysique de Paris, CNRS UMR 7095, Sorbonne
  Universit\'e, 98 bis Bd Arago, 75014 Paris, France}

\date{30 January 2018}

\begin{abstract}
In a companion paper [Pitrou, Phys. Rev. E 97, 043115 (2018)], a
formalism allowing to describe viscous fibers as one-dimensional objects was developed. We apply it to the special
case of a viscous fluid torus. This allows to highlight the
differences with the basic viscous string model and with its viscous
rod model extension. In particular, an elliptic deformation of the
torus section appears because of surface tension effects, and this cannot be described by viscous string nor viscous rod models. Furthermore, we study the
Rayleigh-Plateau instability for periodic deformations around the perfect
torus, and we show that the instability is not sufficient to lead to
the torus breakup in several droplets before it collapses to a single
spherical drop. Conversely, a rotating torus is dynamically attracted
toward a stationary solution, around which the instability can develop
freely and split the torus in multiple droplets.
 \end{abstract}

\maketitle

\section{Introduction}


 
In our companion article~\cite{PitrouPRE1}, we developed a formalism to
describe a viscous fiber as a one-dimensional object with an internal
structure. Given the numerical complexity for solving the full set of
fluid dynamics equations, taking into account junction conditions at
the fiber side for the stress tensor, this description allows for a
simpler route for the numerical resolution of viscous fiber dynamics. 
It is based on an expansion in the slenderness parameter $\epsilon_\ra\equiv \ra/L$, where $\ra$ is the
radius of the fiber and $L$ is the scale associated with typical velocity
gradients. At lowest order in this expansion, the description is
exactly the same as a viscous string, with no internal resistance to
bending nor twisting, and the fiber is only subject to tangential
forces from stretching. However, at the next to leading order, which is ${\cal
  O}(\epsilon_\ra^2)$ smaller, this description departs from the rod
model developed in Refs.~\cite{Fraunhofer1,Ribe2004,Ribe2006}, which is
another theoretical refinement of the viscous string model. The goal
of this article is to emphasize the differences between these two
formalisms, considering a torus of viscous fluid that shrinks due to the
effect of surface tension. After summarizing the formalism in
\S~\ref{SecFormalism}, we show in \S~\ref{SecShrinking} that the shrinking paces between
our model and the rod model are slightly different, and we also emphasize the
importance of the elliptic deformation of the torus section to obtain
a coherent description. In \S~\ref{SecRP} we exhibit the differences in
the Rayleigh-Plateau instability, when considering the dispersion
relation of periodic and linear deformations. We also solve
numerically for the dynamical evolution of these linear
perturbations. Finally, we revisit in \S~\ref{SecRotation} the torus dynamics and the
Rayleigh-Plateau instability for an initially rotating torus.

\section{One-dimensional description}\label{SecFormalism}

In the next section, we briefly review the main features of our one-dimensional model for viscous fibers which is built in
Ref.~\cite{PitrouPRE1}. We then emphasize the differences with the rod model in \S~\ref{SecRod}.

\subsection{Summary of our formalism}

At each time $t$, a viscous fiber is described as a one-dimensional object
through the trajectory $\gr{R}(s,t)$ of its central line, where $s$ is
a length parameter along this central line (see Fig.~1 of
Ref.~\cite{PitrouPRE1} for an illustration). The unit tangent vector to
this central line is
\be
\gr{T} \equiv \partial_s \gr{R}\,
\ee
and the velocity of the central line is
\be
\gr{U} \equiv \partial_t \gr{R}\,.
\ee
Curvature of the central line at a given time is defined as
\be
\gr{\kappa} \equiv \gr{T} \times \partial_s \gr{T}\,.
\ee
A fiber section labeled by a given $s$ is made of points in the viscous fluid which lie in
a plane containing $\gr{R}(s)$ and normal to $\gr{T}(s)$. For each
section we form an orthonormal basis $\gr{d}_\tetr{i} \equiv
(\gr{d}_\tetr{1},\gr{d}_\tetr{2}, \gr{d}_\tetr{3}=\gr{T})$, and by construction the $\gr{d}_\tetr{a},\,a=1,2$,  are tangent to the section. Any vector $\gr{X}$
(e.g. the fluid velocity) can be split into longitudinal components (along $\gr{T}$) and sectional
components (along the $\gr{d}_\tetr{a}$) as
\be\label{BarNotation}
\gr{X} = P_\perp(\gr{X}) + \overline{X} \gr{T}\,,\qquad P_\perp(\gr{X}) \equiv X^a \gr{d}_\tetr{a}\,.
\ee
We also introduce the general notation 
\be
\widetilde{\gr{X}} \equiv \gr{T} \times \gr{X}\,,
\ee
and in particular $\widetilde{\gr{\kappa}}$ is a vector which points toward the exterior of the central line curvature.

The rotation rate of the frame is defined as
\be
\partial_t \gr{d}_\tetr{i} = \gr{\omega} \times \gr{d}_\tetr{i}\,.
\ee
From these definitions, one infers the constraints
\beas\label{EqsConstraints}
(\partial_s \gr{U})\cdot \gr{T}&=&0\,, \slabel{EqsConstraint2}\\
P_\perp(\gr{\omega} )&=& \gr{T} \times \partial_s \gr{U}=\partial_s \widetilde{\gr{U}}+\gr{\kappa}\overline{U} \slabel{EqsConstraint3}\,.
\eeas

The velocity of the fluid $\gr{\totV}$ {\it on the central line} is decomposed into the velocity
{\it of the central line} $\gr{U}$ and the velocity {\it with respect to the central
line} $\gr{V}$, that is as
\be
\gr{\totV} = \gr{U} +\gr{V}\,.
\ee 
In order to obtain a one-dimensional description for the viscous
fiber, we expand in multipoles the variation in each section of the fluid velocity
outside the central line. Several constraints from the boundary
junction condition (which take into account surface tension effects) and from the
Navier-Stokes equation, allow to express all of these multipoles in
functions of $\Sv$ and $\Sw$, which are, respectively, the longitudinal
part of $\gr{V}$ [that is, $\Sv \equiv \overline{V}$ using
notation~(\ref{BarNotation})] and the solid rotation rate of
the fluid around the central line axis $\gr{T}$. Eventually, the
description is reduced to the dynamics of these two
quantities in addition to the central line motion found from the
evolution of $\gr{U}$, combined with the evolution of the section shape which is
allowed to depart from strict circularity.

In this procedure, there is a natural expansion in $\epsilon_\ra$. For instance, terms of the type $\kappa^a
\kappa_a \ra^2 $, where $\ra$ is the fiber radius, are of order ${\cal
  O}(\epsilon_\ra^2)$. Keeping only the lowest order terms amounts to considering the viscous string
model. However, our model for curved fibers in Ref.~\cite{PitrouPRE1}
consists in including the first corrections that are ${\cal
  O}(\epsilon_\ra^2)$. In particular, when including these higher
order effects, the shape of the sections cannot simply be described by
the radius $\ra$ as it deforms. One must also account for an elliptic
deformation, whose dynamics is found from the boundary condition.

\subsection{The rod model}\label{SecRod}

The rod model was used in steady or stationary situations in
Refs.~\cite{Ribe2004,Ribe2006,Fraunhofer1,Fraunhofer2,Fraunhofer3,Fraunhofer4}, but its formulation is general and can account for time dependence. It is based on three balance equations, namely, the matter
balance equation, the momentum balance equation, and the angular
momentum balance equation. We gathered the rod model equations and
a discussion on their shortcomings in \S~ VII.G.4 of Ref.~\cite{PitrouPRE1}. To
summarize, for the rod model one assumes that fiber sections remain
{\it i)} circular, {\it ii)} orthogonal to the fiber central line, and
{\it iii)} that the fluid velocity on the central line is such that
the fluid particle on the central line stays on the central line [$P_\perp(\gr{V})=0$]. 
\begin{itemize}
\item {\it i)} Assuming circular shapes is necessary to obtain
  meaningful momentum and angular momentum balance equations in the
  rod model. However we find that the an elliptic deformation is
  necessarily generated from fiber curvature as an ${\cal
    O}(\epsilon_\ra^2)$ effect. Note that a first attempt of describing elliptic sections, but restricted to straight fibers, was
performed independently in Ref.~\cite{BLF}.
\item {\it ii)} For the fiber section to remain orthogonal to the
  central line, it requires that the solid rotation rate of the section is equal to the rotation rate of the fluid on central
  line at lowest order (see \S~VII.B.4 of Ref.~\cite{PitrouPRE1}). Hence, the angular momentum
  balance cannot be an equation which determines the solid rotation of
  sections since it is determined already by the momentum balance
  equation.  Instead, the sectional viscous forces $P_\perp(\gr{F})$ which lead
  to a net torque $\gr{T} \times \gr{F}$ per unit of fiber length are
  constrained in the rod model by the angular momentum balance so as
  to enforce this property. Since the momentum of inertia of a fiber section scales as
  $\ra^4$ and sectional forces scale as $\ra^2$, we deduce that adding
  sectional forces to the momentum balance equation amounts to
  including order $\epsilon_\ra^2$ terms, which are otherwise absent in the
  lowest order string model.


\item {\it iii)} We showed in \S~IV.I of Ref.~\cite{PitrouPRE1} that to ensure that the central line
  remains on the section center, defined as the position for which
  there is no dipole in the shape deformation, the fluid velocity {\it
    on the central line} ($P_\perp(\gr{\totV})$) must be slightly
  different from the velocity {\it of the central line}
  ($P_\perp(\gr{U})$), that is the difference which is
  $P_\perp(\gr{V})$ does not vanish but is instead of order $\epsilon_\ra^2$. This
  was clearly identified in~\cite{Ribe2006} but ignored precisely on
  the grounds that it was expected to be an order $\epsilon_\ra^2$
  corrections. However, since the inclusion of sectional forces, which
  is the main difference between a rod model and a string model, is
  also a correction of order $\epsilon_\ra^2$, we found that it is
  formally inconsistent to choose one correction and discard another one,
  if they are formally of the same order.
\end{itemize}

\section{Dynamics of a shrinking torus}\label{SecShrinking}

\subsection{Adapted coordinates and variables}

In order to highlight some differences between our formalism and the
rod model, we study the special case of a viscous torus surrounded by
vacuum. The case of a torus surrounded by a highly viscous fluid
has been studied analytically in Ref.~\cite{Yao10} by considering the approximate Stokes flow equation. 

\begin{figure}
\begin{center}
\includegraphics[width=\columnwidth]{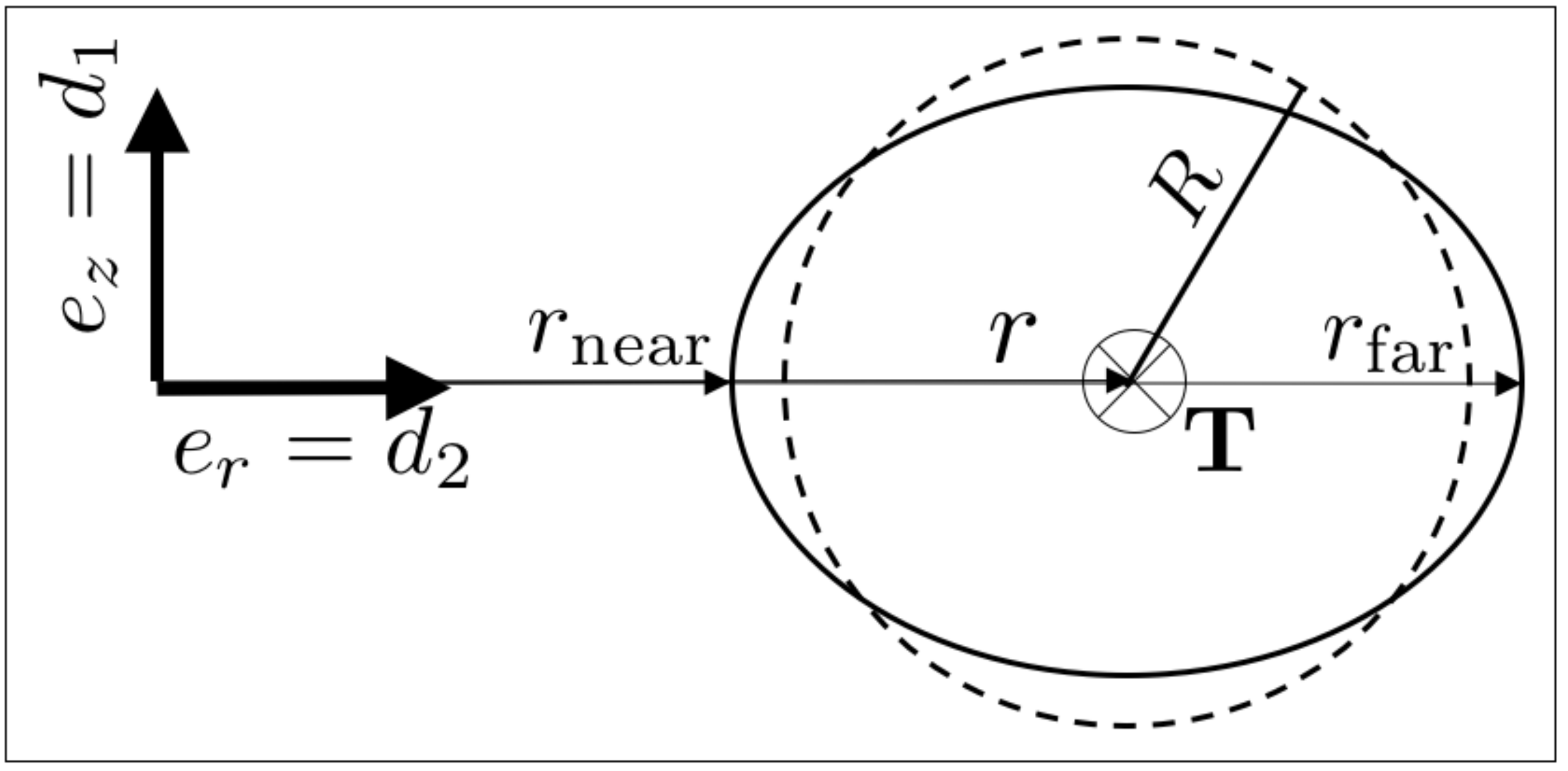}
\caption{Notation in the plane of constant $\theta$.}
\label{FigConstantTheta}
\end{center}
\end{figure}

There are natural cylindric coordinates $(r,\theta,z)$ associated with the torus
with a basis of unit vectors $\gr{e}_r,\gr{e}_\theta,\gr{e}_z$ (see Fig.~\ref{FigConstantTheta} for an illustration of the notation).
Due to its high degree of symmetries, no quantity can depend on
$\theta$, and if the fiber central line is in the plane $z=0$ it
remains so. At any time, the fiber central line is entirely characterized by the radial distance
$r(t)$, hence all quantities can only depend on $t$. In particular,
torus sections cannot mix and there can be no sectional viscous
forces. This implies that the viscous rod model matches exactly the
viscous string model. Hence this system is of particular academic interest because on the contrary our model differs from the string model.

The orthonormal basis associated with the fiber coordinates is
given by 
\be
\gr{d}_\tetr{1} \equiv \gr{e}_z\,,\quad \gr{d}_\tetr{2} \equiv \gr{e}_r\,,\quad \gr{T} \equiv \gr{e}_\theta\,.
\ee
Starting from a reference angle, the relation between the polar angle
$\theta$ and the affine parameter $s$ is simply 
\be\label{stheta}
s = r \theta\,.
\ee
The fiber curvature has only one component $\kappa$ which is 
\be
{\bm \kappa} = \kappa \gr{e}_z\,,\qquad \gr{\tkappa} = \kappa \gr{e}_r \,,\qquad\kappa\equiv
\frac{1}{r}\,.
\ee
From the symmetries, the section ellipticity (see \S~IV.D of
Ref.~\cite{PitrouPRE1} for definitions) is necessarily fully specified
by one polarization $\hrbp$, and it is of the form
\be
\hrb^{ab}\equiv \hrbp (e_z^{\,\,a}e_z^{\,\,b}-e_r^{\,\,a}e_r^{\,\,b})\,.
\ee

The central line sectional velocity is necessarily oriented radially. 
Hence we define
\be
U^a \equiv U^r e_r^{\,a}\,.
\ee
As for the fiber longitudinal velocity $\bar U$, it is not vanishing because the geometric point of constant affine coordinate $s$ has a
longitudinal motion due to the shrinking of the torus. Its value is determined by
the structure relation~(\ref{EqsConstraint2}) which implies
\be
\partial_s \bar U = -\gr{U} \cdot \gr{\tkappa} = -U^a \tkappa_a=-\frac{U^r}{r}\,.
\ee
Since $U^r$ does not depend on $s$, we find 
\be\label{barU}
\bar U = -\frac{s U^r}{r}\,.
\ee
However, the total fluid tangential velocity on the central line is 
\be\label{TotalSv}
\ScV = \bar U+\Sv\,,
\ee 
and from the symmetries of the problem this
quantity does not depend on $s$. If furthermore there is no initial
rotation around the $z$ axis and the fluid falls radially, then
$\ScV=0$. We assume in this section that this is the case and we postpone
the case $\ScV \neq 0$ to \S~\ref{SecRotation}.

One would naively expect that rotation of the fiber central line
$\gr{\omega}$ should vanish due to the symmetries of the
problem. However, it is defined as the rotation of the orthonormal
basis for a geometric point of constant affine coordinate $s$, and it
is instead obtained from the structure relation~(\ref{EqsConstraint3}) which implies
\be
\gr{\omega} = \gr{\kappa} \bar U\,.
\ee 

Finally, the evolution of the radius $\ra$ satisfies
\be\label{EqlnRShrink}
\frac{\partial_t \ra}{\ra} = -\frac{1}{2}\partial_s \Sv = -\frac{U^r}{2 r}=-\frac{\dot r}{2 r}\,.
\ee
This is valid even when including corrections of order
$\epsilon_\ra^2$, and it is an obvious consequence of volume
conservation, which for a torus is $(2\pi r)\times (\pi
\ra^2)$. Hence, one can directly use its first integral which reads as
\be
R=R_i\sqrt{\frac{r_i}{r}}\,.
\ee
Finally, let us define an aspect ratio parameter by
\be
\epsilon_i = \frac{R_i}{r_i}\,,
\ee
which characterizes the initial shape of the torus.

\subsection{Governing equations}

From the dynamical equations gathered in \S~VII.B.9 of
Ref.~\cite{PitrouPRE1} for the lowest order string model and \S~VII.C.7
for the  ${\cal O}(\epsilon_\ra^2)$ corrections, we finally find
\beas\label{SysBackground}
\ddot r= \partial_t U^r &=& -\frac{\ST
  \kappa}{\ra}\left[1+\frac{7}{8}(\kappa \ra)^2\right] \slabel{Masterr2}\\
&& -3 \visc \kappa^2 U^r \left[1+\frac{11}{16}(\kappa \ra)^2\right]\nonumber\\
&&-\frac{21}{16} \kappa^3 \ra^2 (U^r)^2+ \ScV^2\kappa\left[1+ \tfrac{1}{2}(\kappa\ra)^2\right]\,, \nonumber\\
\frac{\dd (\hrbp \ra^2)}{\dd t} &=& -\frac{\ST}{\ra \visc} \left[\hrbp \ra^2
  +\frac{(\kappa \ra)^2}{6}\right] \nonumber\\
&& -\frac{7}{16}\kappa(\kappa \ra)^2 U^r -\frac{1}{12 \visc} (\ScV  \kappa \ra)^2 \,. \slabel{EqRabShrinking}
\eeas
Since we assumed that the fluid is incompressible, we chose a unit mass density $\rho$ so that the viscosity $\visc$ and surface tension
$\ST$ stand in fact for $\visc/\rho$ and $\ST/\rho$. The first line of Eq.~(\ref{Masterr2}) is the radial force from
surface tension which tends to shrink the torus,\footnote{Intuitively
  the extrinsic curvature in the exterior of the torus is larger than
  in the interior, inducing a pressure gradient toward the exterior, hence creating an inward radial force.} with its lowest order
form and its first correction $\propto (\kappa \ra)^2$. The second
line corresponds to viscous friction, similarly given with the lowest
order and its correction $\propto (\kappa \ra)^2$. The first term in the last
line is a convective effect and the last term vanishes if the torus is not rotating initially (see
\S~\ref{SecRotation}) as it corresponds to inertial forces. Eq.~(\ref{EqRabShrinking}) rules the dynamical evolution of
the elliptic deformation. Its last term vanishes if the fluid is not
rotating as it corresponds to tidal inertial forces. In particular, when considered in the limit
of vanishing viscosity, Eq.~(\ref{EqRabShrinking}) amounts to the constraint
\be\label{Qconstraint1}
\hrbp  \ra^2\simeq -\frac{(\kappa \ra)^2}{6}\,.
\ee
It corresponds to the condition for which the quadrupole of extrinsic
curvature [Eq. (7.34c) of Ref.~\cite{PitrouPRE1}] vanishes, inducing no
quadrupole in the pressure distribution inside sections. 

Finally, note that in the inviscid case and still for no rotation
($\ScV=0$), Eq. (\ref{Masterr2}) reduces to
\be\label{Masterr2Inviscid}
\ddot r= -\frac{\ST \kappa}{\ra}\left[1+ \frac{7}{8} (\kappa
\ra)^2\right]-\frac{21}{16} \kappa^3 \ra^2 (\dot r)^2\,.
\ee

\subsection{Analytic approximation}

In this section, we look for approximate analytic solutions using the simpler
viscous string model, which consists in keeping only the dominant term in the first two
lines of Eq.~(\ref{Masterr2}), that is,
\be\label{StringModel}
\ddot r= \partial_t U^r = -\frac{\ST \kappa}{\ra} -3 \visc \kappa^2 U^r\,.
\ee

Let us first consider the inviscid case. The dynamics is simply governed by
\be\label{LowViscositySolution}
\ddot r = -\frac{\ST}{\sqrt{r_i r} R_i}\quad \Rightarrow\quad \dot r =
-2 \sqrt{\frac{\ST(\sqrt{r_i}-\sqrt{r})}{\sqrt{r_i}R_i}}\,.
\ee
It is very similar to a two-body problem in Newtonian gravity with no
initial angular momentum, but with an attractive potential scaling as $\propto
\sqrt{r}$ instead of $\propto-1/r$. It is further integrated as
\be\label{Implicitr}
1 - \tfrac{3}{4} x - \tfrac{1}{4}x^{3/2}=\frac{\ST}{R_i r_i^2} \left(\frac{3t}{4}\right)^2\,,\qquad
x\equiv \frac{r}{r_i}\,,
\ee
which needs to be solved algebraically to obtain $r(t)$. In particular
when $r$ remains close to $r_i$, that is, for the beginning of the
torus shrinking, we obtain the parabolic motion
\be\label{EqDarbois}
r \simeq r_i -\frac{\ST}{2 r_i R_i} t^2\,,
\ee
which matches the approximate solution (3.6) of
Ref.~\cite{Dar13} in the limit $R_i \ll r_i$. However, since this is only an approximation implicit solution to
Eq. (\ref{Implicitr}), it also tends to overestimates $r$ at late times,
and this can be checked visually on Fig.~4 of Ref.~\cite{Dar13}. Indeed, since
Eq.~(\ref{EqDarbois}) is the solution of  $\ddot r = -\ST/(r_i R_i)$, it
underestimates the inward acceleration when compared to
Eq. (\ref{LowViscositySolution}). We use Eq.~(\ref{EqDarbois}) to
define the collapse time scale for the inviscid case (the capillary
time scale) as
\be\label{tcap}
t_{\rm cap} \equiv \frac{r_i \sqrt{R_i}}{\sqrt{\nu}}\,.
\ee

For high-viscosity cases, we need to solve instead the quasistatic
approximation ($\ddot r \simeq 0$), equivalent to the Stokes flow equation, which leads to
\be
\frac{3 \visc \dot r}{r^2} = -\frac{\ST}{r R} = -\frac{\ST}{R_i
  \sqrt{r_i r}}\,.
\ee
Its solution is
\be
t = \frac{6 \visc R_i}{\ST} \left(\sqrt{\frac{r_i}{r}}-1\right)\,,\quad
r = \frac{r_i}{\left(1+\frac{\ST t}{6 \visc R_i}\right)^2}\,.
\ee
Hence, for high-viscosity, the collapse time scale is
\be
t_{\rm visc} \equiv \frac{6\mu R_i}{\nu}\,.
\ee
Viscosity can be neglected when $t_{\rm visc} \ll t_{\rm cap}$, that is, if
\be
\visc \ll \visc_0\equiv \frac{r_i}{6}\sqrt{\frac{\ST}{R_i}} =\frac{1}{6}\sqrt{\frac{\ST r_i}{\epsilon_i}}\,.
\ee

\subsection{Numerical resolution}

\begin{figure}[!htb]
\includegraphics[width=\columnwidth]{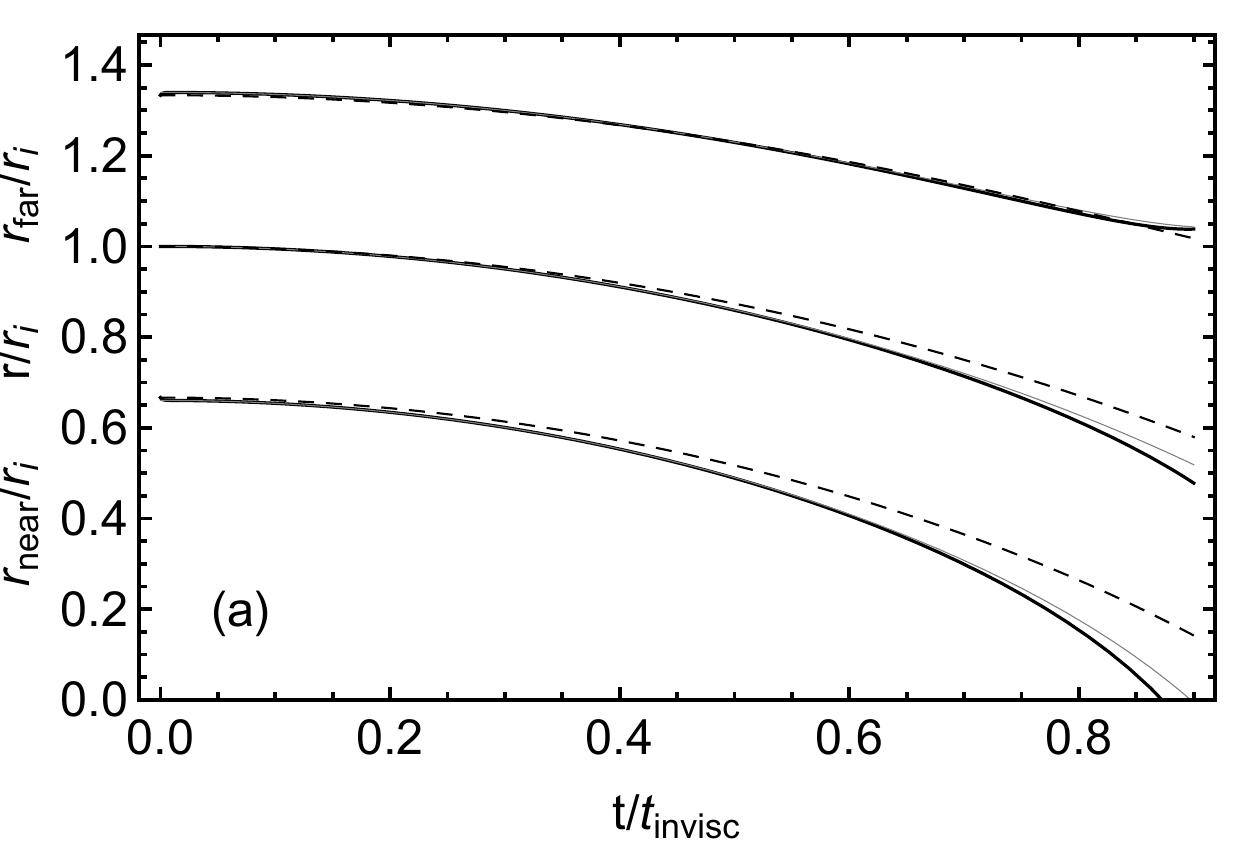}
\includegraphics[width=\columnwidth]{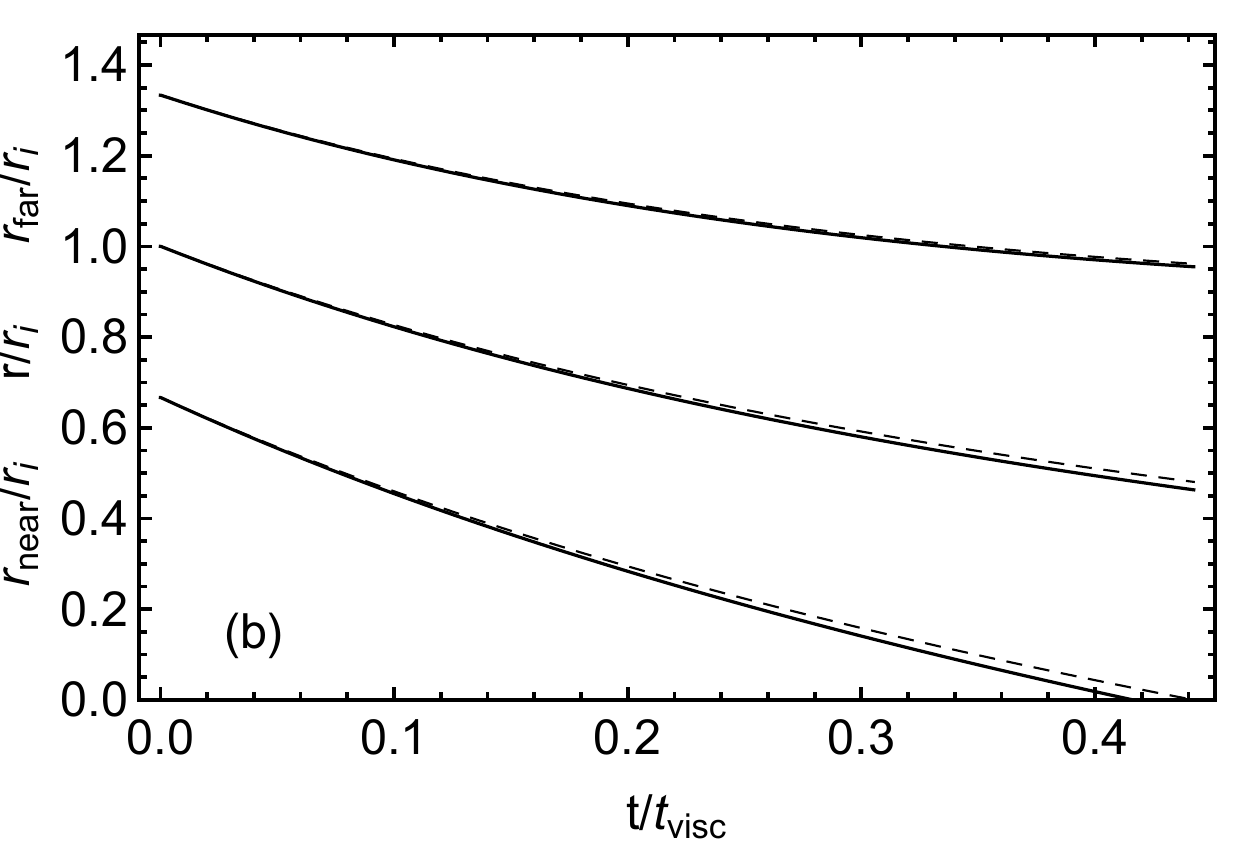}
\caption{The initial torus shape is $\epsilon_i = 1/3$ in both
  cases. Fig. (a) is a low viscosity case ($\mu=0.01 \mu_0$)
  and Fig. (b) is a high viscosity case ($\mu = 100 \mu_0$). The continuous
  lines correspond to our model, the dashed lines to the basic rod
  model which is equivalent to the viscous string model in this case. The central curves correspond to the central
line, the top curves to the outer intersection of the section with the
torus plane ($z=0$), and the lower curves to the inner
intersection. 
In Fig. (a), the inviscid solution [Eq.~(\ref{Masterr2Inviscid})] is depicted in a thin and continuous gray line.}
\label{Fig1}
\end{figure}

The numerical resolution of the full system of Eqs.~(\ref{Masterr2}) and
(\ref{EqRabShrinking}) is illustrated in Fig.~\ref{Fig1} for typical high and
low viscosities.  In the low-viscosity case, we also plot the
solution to the inviscid equation~(\ref{Masterr2Inviscid}). Given the
elliptical deformation of the torus sections, the far and near sides of the section are located respectively at 
\bea
r_{\rm far} &=& r + R(1-\hrbp \ra^2)\,\\
r_{\rm near}  &=& r - R(1-\hrbp \ra^2)\,.
\eea

We observe that in all cases, the rod model underestimates the shrinking of the central
line. The numerical results are in agreement with our expectation that the difference
between the two descriptions should be of order of the initial
$\epsilon_\ra^2=(\kappa \ra)^2=(\ra/r)^2$ which we have taken to be
$(1/3)^2$ in initial conditions.
Furthermore, the elliptic deformation reduces even further the
distance of the inner intersection of the section with the torus plane
($z=0$) since numerically we get $r_{\rm near} < r-R $. The torus is deformed as if it was squeezed in the azimuthal
direction, or as if it was stretched by radial tidal forces. When the inner point reaches a null radius, that is
approximately when $r=\ra$, the expansion is expected to break down
since the ratio $\kappa \ra=\ra/r$ reaches unity. Physically it also
corresponds to the point where the torus becomes topologically a
sphere and an accurate description of the final ringdown toward a
sphere must be performed from a multipolar expansion around such a
geometry, as e.g. in Ref.~\cite{Cha59} for linear dynamics or in
Ref.~\cite{Bec94} for non-linear dynamics.

\section{Rayleigh-Plateau instability}\label{SecRP}

\subsection{Straight fibers}

It is well known that periodic radius perturbations around an infinite
straight fiber are unstable for $k \ra < 1$, where $k=2\pi/\lambda$ is the mode of
the perturbation. This is the celebrated Rayleigh-Plateau (RP) instability
\cite{Rayleigh1878,Plateau1873,Rayleigh1892,Weber1931} (see also
Refs.~\cite{EggersRMP,Eggers2008} for reviews). In the case of an
elongated but finite fiber, the RP has been studied numerically in Ref.~\cite{Dri13}. 
The RP instability is understood analytically by considering periodic radius perturbations
around an infinite cylinder of viscous fluid with radius $\ra_0$ of the form
\be\label{Rexpansion}
\ra = \ra_0 + \delta \ra\,,\qquad \delta \ra = \underline{\delta \ra} {\rm e}^{\alpha t}
\cos(k s),
\ee
and then linearizing the Navier-Stokes equation and the boundary
constraint. Using units for which $\ra_0 = 1$ and $\ST=1$ for convenience, this leads to the implicit dispersion relation~\cite{Weber1931,GarciaCastellanos}
\bea\label{RayleighImplicit}
&&\alpha^2 \frac{k I_0(k)}{2 I_1(k)} =\tfrac{1}{2}(k^2-k^4)\\
&&- \visc \alpha k^2 \left[\frac{2k
    I_0(k)}{I_1(k)}-1+\frac{2 \visc k^2}{\alpha}\left(\frac{k
      I_0(k)}{I_1(k)}-\frac{k_1 I_0(k_1)}{I_1(k_1)}\right)\right]\nonumber
\eea
where the $I_n$ are the Bessel's functions and with
\be
k_1^2 \equiv k^2+\frac{\alpha}{\visc}\,.
\ee
It is then found that in the inviscid case, the mode which grows the most rapidly
corresponds to $k_{\rm max} R_0 \simeq 0.69702$, that is
\be\label{kmax}
\lambda_{\rm max} = \frac{2\pi}{k_{\rm max}} \simeq 9.0144 R_0\,.
\ee 
Since Eq.~(\ref{RayleighImplicit}) is implicit, it is convenient to
expand it in powers of $k$ (still in units where $R_0=\ST=1$), and we
get [see Eq. (96) of Ref. \cite{GarciaCastellanos}]
\bea\label{AlphaRayleigh}
\alpha &\simeq& \frac{k}{\sqrt{2}}-\frac{3\visc}{2}k^2+\frac{36
  \visc^2-9}{16\sqrt{2}}k^3 +\frac{3 \visc}{16} k^4 \\
&+&\frac{-49+360\visc^2-1296
  \visc^4}{512\sqrt{2}}k^5+\left( \frac{1}{3072 \visc} -\frac{\visc}{96}\right)k^6\,.\nonumber
\eea
Note that this series is not valid in the limit $\visc \to 0$. However, it allows
to compare the exact result (\ref{RayleighImplicit}) with the one obtained from the
straight fiber results found with our method in \S~VI of Ref.~\cite{PitrouPRE1}. Eqs.~(6.10a) for $\partial_t \Sv$ [with its ${\cal O}(\epsilon_\ra^2)$ and ${\cal
  O}(\epsilon_\ra^4)$ corrections given in Eqs.~(6.23a) and (E1)],
together with Eq.~(6.3) for $\partial_t \ra$  [with its ${\cal O}(\epsilon_\ra^2)$ and ${\cal
  O}(\epsilon_\ra^4)$ corrections given in Eqs.~(6.23c) and (E2)],
once linearized lead to (still in units where $\ST=R_0=1$)
\beas
\partial_t \Sv&=&-\partial_s \delta {\cal K}+3\visc \partial_s^2 \Sv +
\tfrac{3}{8}\visc \partial_s^4 \Sv+ \tfrac{1}{48}\visc \partial_s^6
\Sv\nonumber\\
&&-\tfrac{1}{48 \visc}\partial_t \partial^3_s \delta {\cal
  K}+\tfrac{3}{64}\partial_s^5 \delta {\cal K}\,,\\
\partial_t \delta R &=& -\tfrac{1}{2}\partial_s \Sv
-\tfrac{1}{16}\partial^3_s \Sv +\tfrac{1}{128}\partial^5_s
\Sv\nonumber\\
&&-\tfrac{1}{96 \visc}\partial_s^4 \delta{\cal K}\,,\\
\delta {\cal K} & =& -\delta R -\partial_s^2 \delta R\,.
\eeas
From these we can also determine a dispersion relation. It is achieved
by expanding $\delta \ra$ and  $\partial_s \Sv$ as in
Eq.~(\ref{Rexpansion}). Developing the dispersion relation obtained in powers of
$k$, we get
\bea\label{AlphaMoi}
\alpha &\simeq & \frac{k}{\sqrt{2}}-\frac{3\visc}{2}k^2+\frac{36
  \visc^2-9}{16\sqrt{2}}k^3 +\frac{3 \visc}{16} k^4 \\
&+&\frac{-49+360\visc^2-1296
  \visc^4}{512\sqrt{2}}k^5+\left( \frac{1}{1536 \visc} -\frac{\visc}{96}\right)k^6\,.\nonumber
\eea
One notices that it differs from Eq.~(\ref{AlphaRayleigh}) only in the terms which are $\propto
1/\visc$. This is because our formalism is based on the fact that the
fluid has non-vanishing viscosity and is ill defined in the inviscid
case. However, we note that cancellations occur for the lowest order and the ${\cal O}(\epsilon_\ra^2)$ corrections, and these divergent
terms occur only when including ${\cal O}(\epsilon_\ra^4)$ corrections,
which contribute only to the $k^5$ and $k^6$ terms of the expansion and
beyond.

\subsection{Instability around torus}\label{SecRPTorus}

We now study how this instability is modified when considering small
linear fluctuations around the shrinking torus described previously in
\S~\ref{SecShrinking}. In the case where the torus is surrounded by a
more viscous fluid than the one in the torus, this was already studied
experimentally in Ref.~\cite{Pai09} by forming a torus thanks to
extrusion through a needle, or in Ref.~\cite{McGraw15} using a glassy
transition from solid phase to viscous phase in a polystyrene torus. In both
cases, it is found when $R_i \ll r_i$ that the final state is a breakup of the initial torus into several droplets. This is confirmed by
the numerical study of Ref.~\cite{Meh13}, which considers quadrupolar
perturbations. However, in our case, the torus is not surrounded by a highly viscous
fluid which retards its shrinking, hence the dynamics of the RP
instability is bound to be different. Our system is in fact very similar
to the experiment described in Ref.~\cite{Dar13} where liquid oxygen is sculpted
into a torus thanks to its paramagnetic property, and it differs
essentially only in that we have not included gravity. As it levitates one
its own vapor, this torus is not subject to external viscous
forces. It is found experimentally in Ref.~\cite{Dar13} that the RP instability is never
strong enough to breakup the torus in multiple droplets before its
shrinking reduces it topologically to a sphere, a result that we confirm with our model in this section.

\begin{figure*}[!htb]
\includegraphics[width=0.47\linewidth]{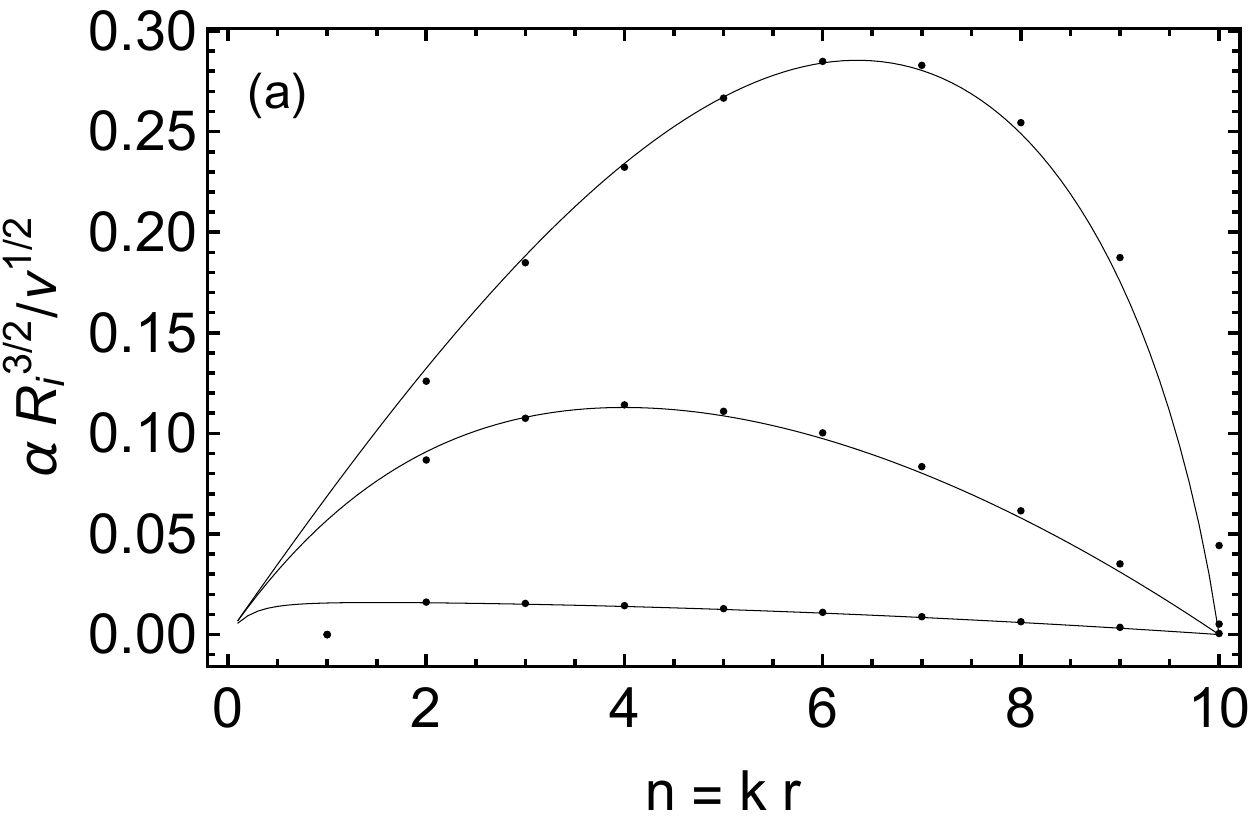}
\includegraphics[width=0.47\linewidth]{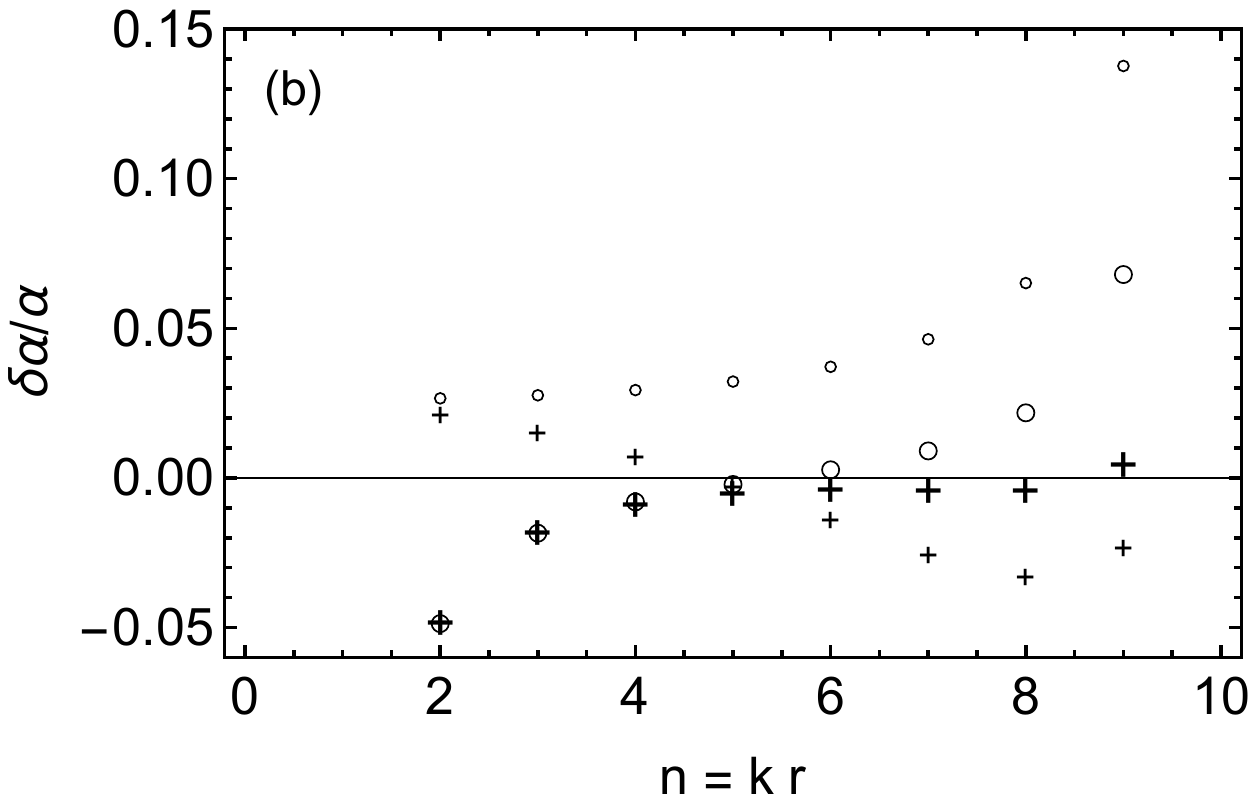}
\caption{The initial aspect ratio is $\epsilon_i=0.1$. Fig. (a) : The dots are the adimensionalized growth rate
  $\alpha R_i^{3/2}\ST^{-1/2}$ as a function
  of the mode integer $n$, for an initial shape characterized by $\epsilon_i=0.1$. The
  continuous line corresponds to the result obtained for the
  Rayleigh-Plateau instability on an infinite straight line. From top
  to bottom we show the cases $\mu = 0.1{\sqrt{\ST R_i}}$,  $\mu =
  {\sqrt{\ST R_i}}$, and  $\mu = 10{\sqrt{\ST R_i}}$. \\
Fig. (b) : Growth rate relative difference with respect to the one obtained for
  the Rayleigh-Plateau instability of an infinite straight line. Our model (circles) and the
  improved rod model (crosses) is evaluated for large symbols with
  $\mu = 0.1\sqrt{\ST R_i} $ (low viscosity) and for small symbols with $\mu = 10^3
  \sqrt{\ST R_i} $ (high viscosity). In all cases, the dipolar mode ($n=1$) growth rate is identically vanishing
  as required by center of mass conservation.}
\label{Fig2}
\end{figure*}

When restricting to linear fluctuations around the shrinking torus (that we consider as a background), great
simplifications arise because the background quantities, that we note
in this section, $\ra_0$, $\kappa_0$, $r_0$, $\hrbp_0$ and $U^r_0$, do not depend\footnote{$\bar U_0$ is given by Eq. (\ref{barU}), and depends
on $s$, but the relevant quantity for the fluid tangential velocity is
$\ScV_0=\bar U_0 + \Sv_0$ which does not depend on $s$.}  on
$\theta$. Hence, we can ignore small angular displacements of the fiber sections due to the
fluctuations. In practice, everything happens as if a section which is
initially at a given $\theta = s/r$ remains at the same angular
position even though $s$ and $r$ vary because of shrinking, indicating
that the variables $(t,\theta)$ are more adapted for the problem than $(t,s)$. 
Note that small angular displacements would only be relevant if we were to consider the non-linear dynamics. 

We split variables into their background and their perturbed quantity as 
\bea
\kappa &=& \kappa_0 + \delta \kappa\,,\\
r&=& r_0 + \delta r\,,\\
\hrbp &=& \hrbp_0 + \delta \hrbp\,,\\
U^r &=& U^r_0 + \delta U^r\,,\\
\ra&=& \ra_0 + \delta \ra\,,\\
\ScV &=& \ScV_0 +  \delta \ScV\,.
\eea
The dynamics of the background has already been discussed in
\S~\ref{SecShrinking} and, considering no torus rotation ($\ScV_0=0$), it is governed by the differential system
(\ref{SysBackground}) in the variables $(r_0, \hrbp_0)$.

For small linear perturbations, we can use the relations
\bea\label{dkappadUr}
\delta \kappa &\simeq& - \frac{\delta r}{r^2}-\partial^2_s \delta r\,,\\
\delta U^r &\simeq& \partial_t \delta r|_\theta = \partial_t \delta r|_s- \bar U_0 \partial_s \delta r\,,
\eea
so that eventually the perturbed set  of variables is 
\be\label{PerturbedVariables}
(\delta r, \delta \ra, \delta \ScV, \delta \hrbp).
\ee 
The linearized equations of our model are reported in the Appendix and one needs only
to replace the relations (\ref{dkappadUr}) to get a closed system of differential equations in these variables.

We first study the growing modes of the Rayleigh-Plateau instability
by considering these equations at initial time, that is, ignoring the
effects of the secular background shrinking. When considering periodic
perturbation, the variables (\ref{PerturbedVariables}) are expanded as
\be\label{DefLinear}
x = \underline x {\rm e}^{\alpha t} \cos(n \theta)=  \underline x {\rm e}^{\alpha t}\cos\left[\frac{n s}{r(t)}\right]\,.
\ee
Solving for $\alpha$, we obtain the dispersion relation as a function of the mode
$n$. The algebraic expressions for $\alpha$ are far too complex to be
reported explicitly. Their numerical values are, however, depicted in the left plot of Fig.~\ref{Fig2}
where we superimposed the Rayleigh-Plateau dispersion relation
(\ref{RayleighImplicit}) obtained in the straight fiber case. The relative differences for our model
and for the improved rod model  [see Eq.~(7.74) of
Ref. \cite{PitrouPRE1}] are presented in the right plot of
Fig.~\ref{Fig2}, and these differences come from the fact that the
models differ in their ${\cal O}(\epsilon_\ra^2)$ corrections.

\begin{figure*}[!htb]
\includegraphics[width=0.47\linewidth]{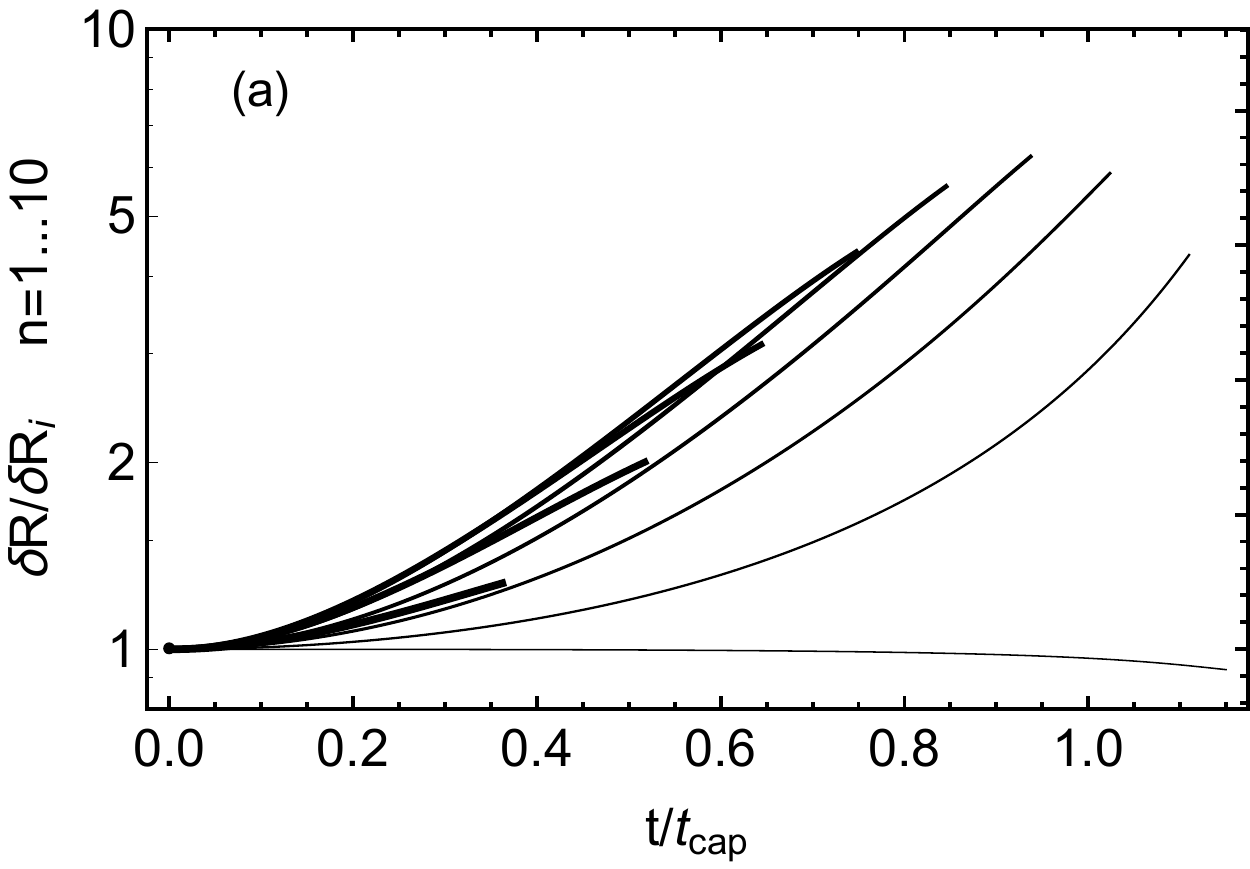}
\includegraphics[width=0.47\linewidth]{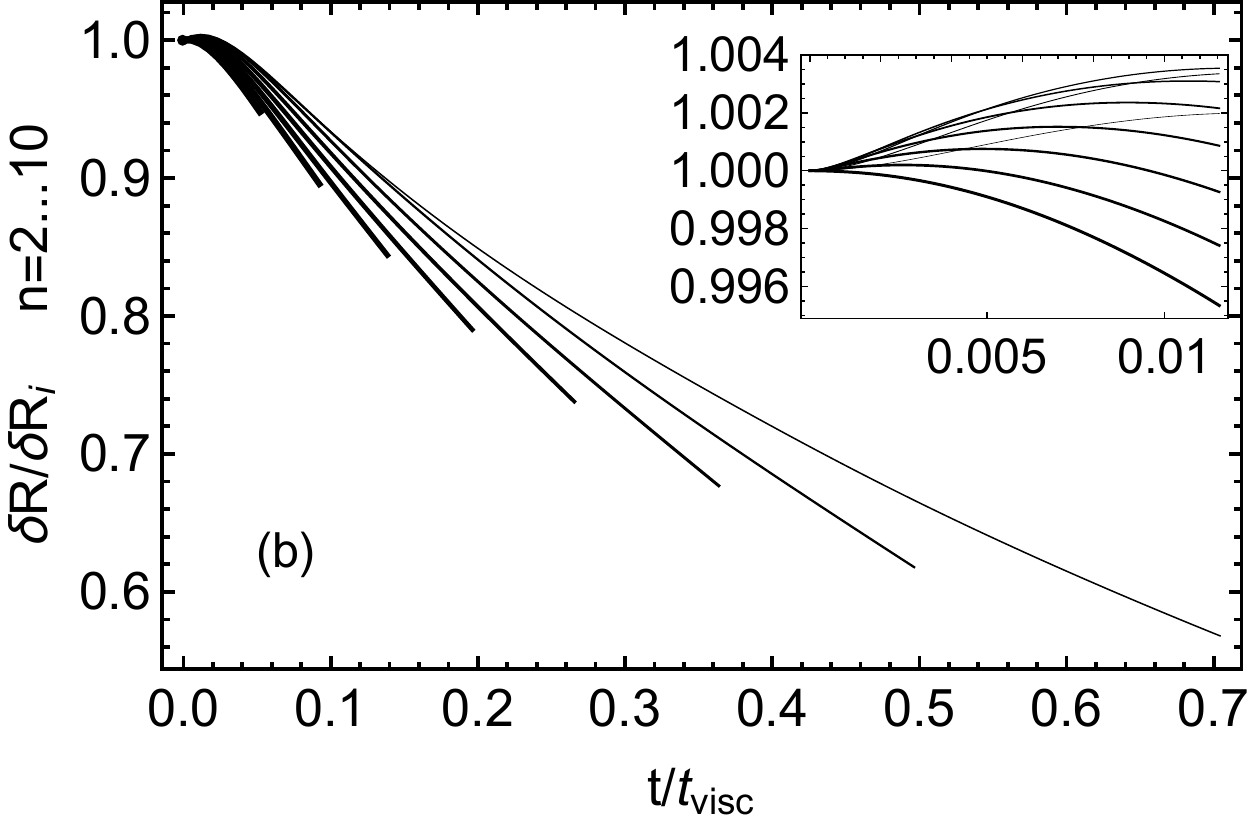}\\
\includegraphics[width=0.47\linewidth]{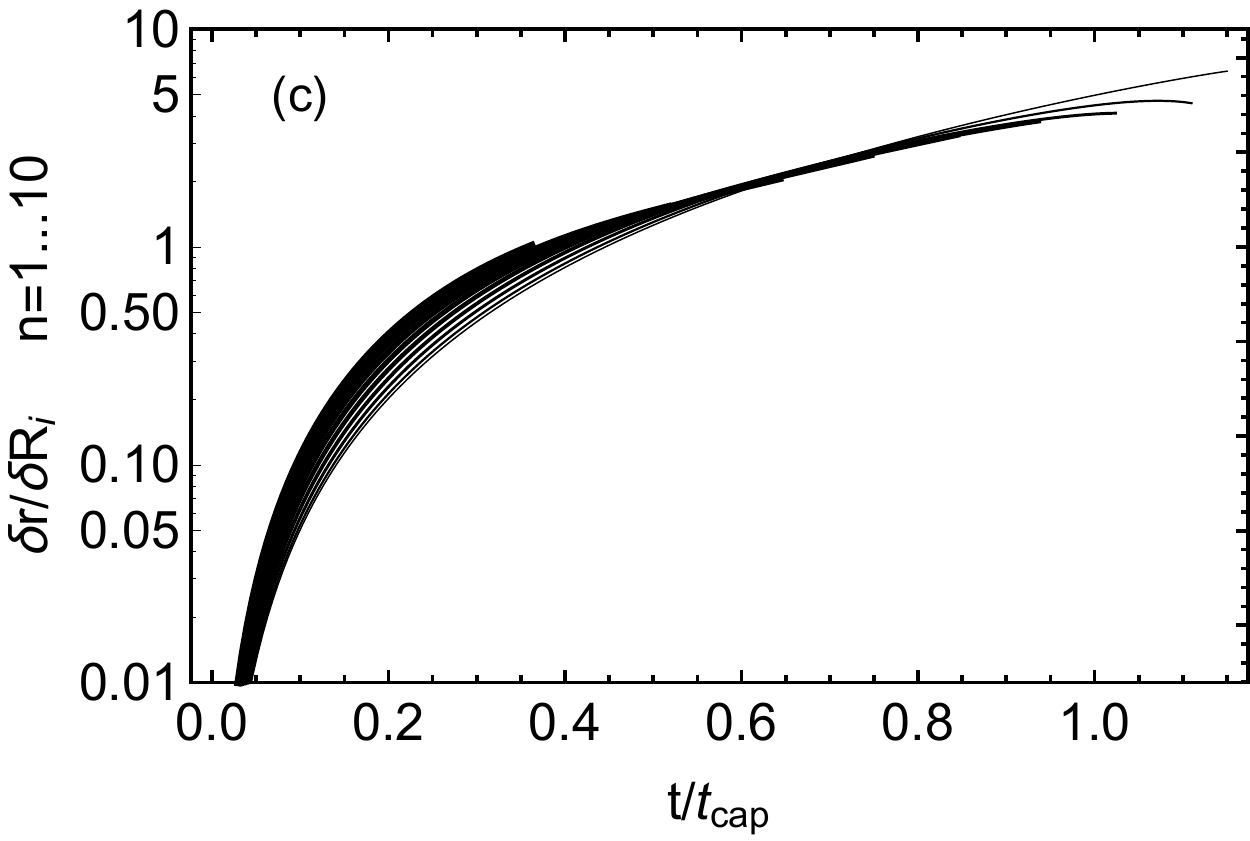}
\includegraphics[width=0.47\linewidth]{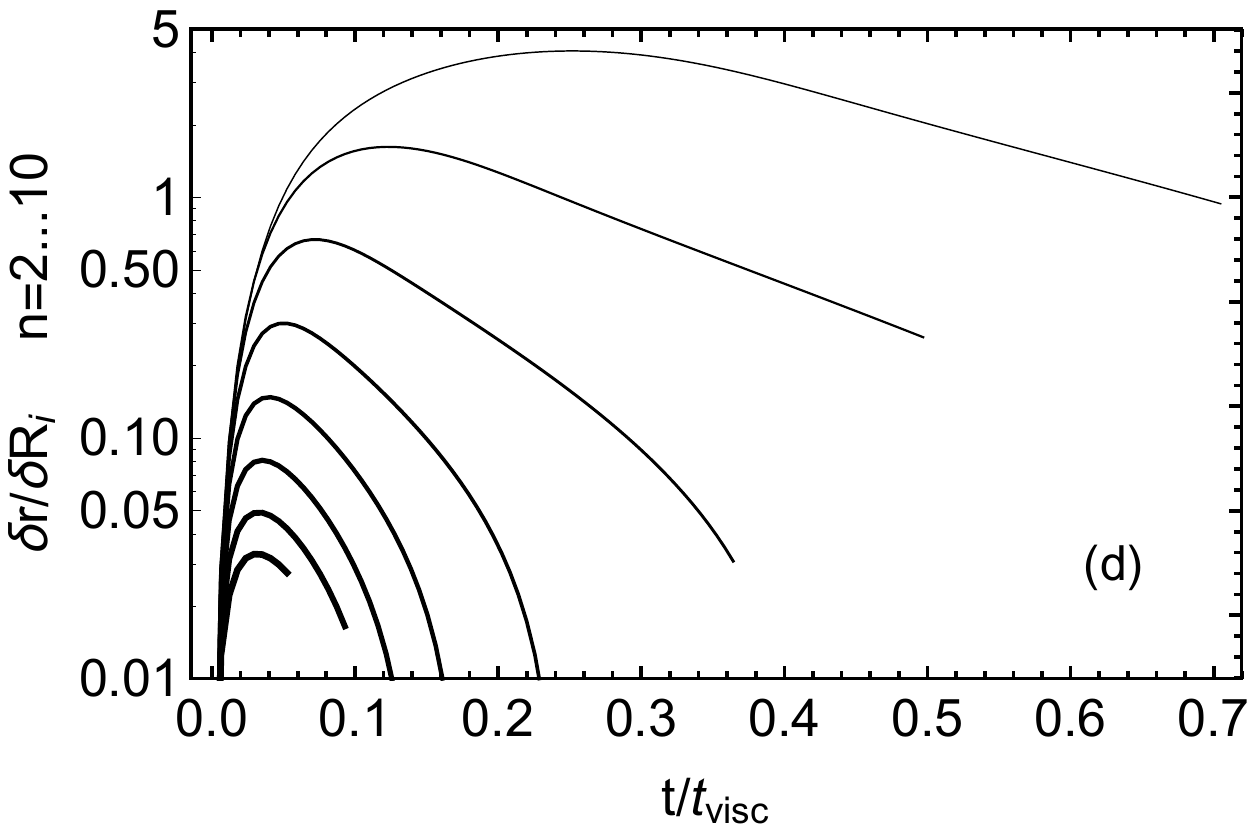}
\caption{(a), (b) : $\delta R/\delta R_i$. (c), (d) : $\delta r/\delta R_i$. The initial
  shape is characterized by $\epsilon_i=0.1$. The viscosity is low ($\mu = 0.1
  \mu_0$) for (a) and (c) and high ($\mu = 10 \mu_0$) for (b) and (d). The instability modes are plotted  from $n=1$ to $10$ for  (a) and (c), or from $n=2$  to $10$ for (b) and (d), starting from the thinnest line to the thickest line.}
\label{Fig3}
\end{figure*}

We then consider the full system of differential equations in
the Appendix and solve for it numerically for different
viscosities. We assume that initially the perturbation is only a
perturbation in the section radius $\delta R_i$, with all other
perturbations vanishing. The growth of the linear perturbations is
then characterized by $\underline{\delta R}/\underline{\delta R_i}$,
which characterizes the RP instability, and which is plotted in Figs.~\ref{Fig3}(a)
and \ref{Fig3}(b). Additionally, departure from the toroidal  geometry of
the central line is characterized by the evolution of $\delta
r$. Since we consider linear perturbations, its evolution is
commensurate with $\delta R_i$, hence we plot in Figs.~\ref{Fig3}(c)
and \ref{Fig3}(d) $\underline{\delta r}/\underline{\delta R_i}$
that indicates how efficiently a deformation of the sections results in a
deformation of the central line. The integration for a
given mode $n$ is stopped at $t_{\rm end}$ defined by the condition $k R = n R /r = 1 $, since it corresponds to
$\epsilon_\ra =1$, that is, to a breaking of the perturbative expansion
in the fiber radius, which is central to the construction of our
model. This happens unavoidably because $r$ decreases as the torus shrinks, but also $R$ increases from volume
conservation. 

\begin{itemize}
\item For low viscosities, slightly before $t_{\rm end}$, we see an
  inflexion point in the perturbation growth. In the light of the RP instability for
straight fibers, and considering that modifications brought by the
toroidal shape are small, this corresponds to the fact that when $k =
n/r > k_{\rm max}$, the value of $\alpha$ in the dispersion relation
decreases to reach $\alpha=0$ when $k R =1$. For each mode, everything happens as if the mode was
climbing the dispersion relation curve $\alpha(k)$ of the usual
straight fiber RP instability from low values of
$k$, to high values of $k$, hence passing through the maximum value
$\alpha(k_{\rm max})$. And in that process, long modes which
correspond to low values of $n$ (but not the dipolar perturbation
corresponding to $n=1$ which is otherwise constrained by center of
mass conservation) have had more time to grow so as to reach the
highest growth. However, we notice that even for the long modes, the
typical growth never reaches huge values. This is because the time
scale of the RP instability is $t_{\rm RP} \equiv \sqrt{R_i^3/\ST}$
and it is related to the collapse time scale (\ref{tcap}) by
$t_{\rm RP} = \epsilon_i t_{\rm cap}$. The RP instability requires a very
  thin torus ($\epsilon_i \ll 1$) to be able to break it before it has
  collapsed. Hence, for reasonable values of $\epsilon_i$, the RP
  instability cannot destabilize the toroidal shape and split in
  several droplets, in agreement with the findings of
  Ref.~\cite{Dar13}. Hence, for the low viscosity case, the main features of the dynamics can be understood from the perspective
of the RP instability around straight fibers. Note that if shrinking is slowed by an
external viscous fluid, it is found experimentally \cite{Pai09,McGraw15} that these conclusions are reversed.

\item For larger viscosities, $k_{\rm max}$ corresponding to the fastest
  growing mode of RP instability around a straight fiber is much
  smaller. Indeed, it scales approximately as~\cite{Dri13} $k_{\rm max} \propto
  1/\sqrt{2+3\sqrt{2} \mu}$ (still in units where $R_0=\ST=1$).
Furthermore, the corresponding value $\alpha(k_{\rm max})$ is much
smaller as illustrated in Fig. \ref{Fig2} (left plot). Hence, if we
were to guess the behaviour from the analogous RP instability around
straight lines, one would conclude that perturbations do not grow
significantly. In fact, we find numerically that perturbations grow
indeed mildly, but only up to around $t_{\rm cap}$, as they are damped afterwards. In that case, there is no RP instability and the
final state is a single drop.
\end{itemize}

\section{Rotating torus}\label{SecRotation}

It is possible to prevent the torus from shrinking by considering a torus
rotating around the $z$ axis. From angular momentum conservation, this
brings a potential barrier and if initial rotation is strong enough, it
prevents the collapse of the torus. Indeed, the last term of
Eq. (\ref{Masterr2}) acts as a repulsive force. 
The dynamics of $\ScV$, which is not identically vanishing anymore, is ruled by
\be\label{Modif2}
\frac{\dd \ScV}{\dd t} = -\ScV U^r \kappa \left[1-\tfrac{9}{4}(\kappa
\ra)^2\right]\,.
\ee
When considering only the leading term in this equation, it can be put
in the form $\dd(r \ScV)/\dd t =0$ which is obviously angular momentum
conservation,\footnote{Angular momentum per unit of mass is $r \ScV[1+3
  R_i^2 r_i/(4 r^3)]$ and Eq. (\ref{Modif2}) implies its
  conservation up to ${\cal O}(\epsilon_\ra^4)$ corrections.} that is,
\be
r \ScV = r_i \ScV_i\,.
\ee 
When the torus shrinks ($U^r <0$), $\ScV$ must
increase. As in the two-body problem of Newtonian gravity, this leads
to a repulsive potential $\propto 1/r^2$, that is a radial force
$\propto 1/r^3$ given by the last term of Eq. (\ref{SysBackground}).

\begin{figure*}[!htb]
\includegraphics[width=0.47\linewidth]{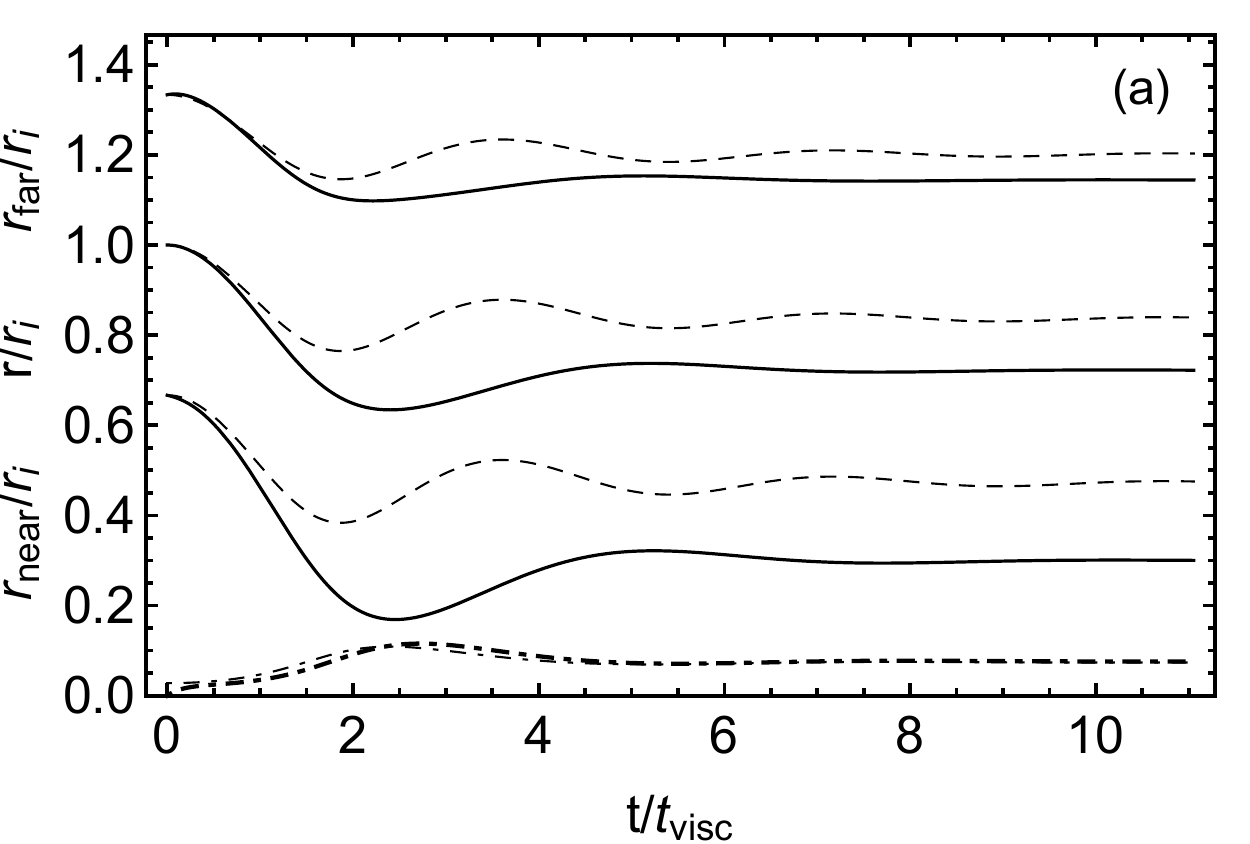}
\includegraphics[width=0.47\linewidth]{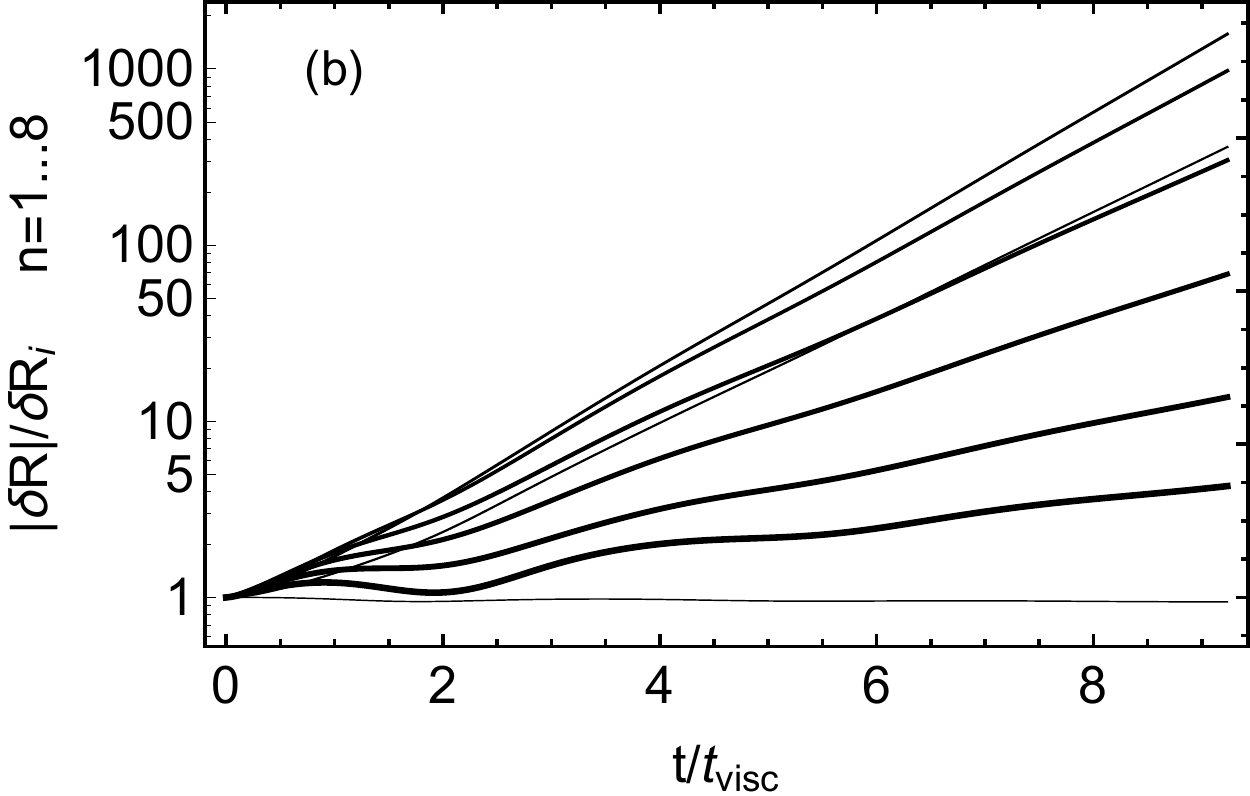}
\caption{For both figures $\ScV_i =0.8 \sqrt{\ST/R_i} $, $\mu =
  \mu_0$. Fig. (a) :  $\epsilon_i = 1/3$. The continuous lines correspond to our
  model and the dashed lines correspond to the viscous string
  model with circular sections. The thick dotted-dashed line is $-\hrbp \ra^2$ whereas the thin
  dotted-dashed line is $R^2/(4 r^2)$. Fig. (b) : $\epsilon_i =0.1$ and
  the lines (from thin to thick) correspond to the modes from $n=1$ to
  $8$ of $|\delta R|/\delta R_i$. The dipolar mode ($n=1$) does not grow because of center of mass conservation.}
\label{FigRotation}
\end{figure*}

If rotation is strong enough, the radius $r$ undergoes damped oscillations (except in the pure inviscid case) to
reach an equilibrium value. If we ignore the ${\cal
  O}(\epsilon_\ra^2)$ corrections, that is, considering only the
viscous string model, this is the position for which surface tension
attractive force is balanced by the repulsive inertial force, and we find
\be\label{Equilibriumr}
r_{\rm stable} \simeq r_i \left(\frac{R_i \ScV_i^2}{\ST} \right)^{2/5}\,.
\ee
Hence, by choosing $\ScV_i = \sqrt{\ST/R_i}$ the torus is directly [up
to the  ${\cal O}(\epsilon_\ra^2)$ corrections] in its stable position.

Furthermore, now the last term of Eq. (\ref{EqRabShrinking})
contributes and it corresponds to the deformation induced by tidal forces. In the inviscid limit, the constraint (\ref{Qconstraint1}) becomes
\be
\hrbp \ra^2 \simeq- \frac{(\kappa \ra)^2}{6}\left(1+\frac{\ScV^2 \ra}{2 \ST}\right)\,.
\ee
Once the radius has reached its equilibrium value
(\ref{Equilibriumr}), the elliptic deformation tends to $\hrbp \ra^2
\simeq -R^2/(4r^2)$ as illustrated in the left plot of
Fig.~\ref{FigRotation}. We then repeat the linear perturbation
analysis of \S~\ref{SecRPTorus}, but including all  the terms involving $\ScV_0$ (which we
do not report in the Appendix), whose dynamics is given by
Eq.~(\ref{Modif2}). Instead of expanding the linear perturbations with
$\cos(ns/r)$ as in Eq. (\ref{DefLinear}), we expand them with
$\exp(\ii ns/r)$. Indeed, the amplitude $\underline x$ becomes complex
when including torus rotation. Its norm still characterizes the
instability, that is the size of the perturbation, and its phase
corresponds to the angular rotation induced by advection. In the right
plot of Fig.~\ref{FigRotation}, we illustrate how the RP instability
can develop freely once the torus has reached a stable rotating
solution. Hence it is expected that the initially rotating viscous
torus will break up in several droplets that will be ejected outward.

\section{Conclusion}

The viscous torus allows to clearly emphasize the difference between our model and the viscous
string or its refined rod model, since for the main shrinking
behaviour these have no corrections of order ${\cal
  O}(\epsilon_\ra^2)$ whereas our model does. Our corrections affect the central
line dynamics, and we find it necessary to also describe the elliptic
deformations. When studying the Rayleigh-Plateau instability, our
model and the (improved) rod model also differ slightly as shown by
considering the dispersion relation. When the torus is not surrounded
by a viscous fluid, our linear analysis indicates that the torus does
not break up into small droplets for all possible viscosities, since
either the instability does not develop in very viscous fluids or does
not have sufficient time to develop in low-viscosity fluids. However,
if the torus is rotating around its geometric center fast enough, it can reach an equilibrium configuration around which the RP instability
should lead to its unavoidable breakup in several droplets.

\begin{acknowledgments}

I thank the anonymous referee of~\cite{PitrouPRE1} for suggesting the toroidal
geometry as a practical application.
\end{acknowledgments}

\bibliography{BiblioJets}

\begin{thebibliography}{23}
\expandafter\ifx\csname natexlab\endcsname\relax\def\natexlab#1{#1}\fi
\expandafter\ifx\csname bibnamefont\endcsname\relax
  \def\bibnamefont#1{#1}\fi
\expandafter\ifx\csname bibfnamefont\endcsname\relax
  \def\bibfnamefont#1{#1}\fi
\expandafter\ifx\csname citenamefont\endcsname\relax
  \def\citenamefont#1{#1}\fi
\expandafter\ifx\csname url\endcsname\relax
  \def\url#1{\texttt{#1}}\fi
\expandafter\ifx\csname urlprefix\endcsname\relax\def\urlprefix{URL }\fi
\providecommand{\bibinfo}[2]{#2}
\providecommand{\eprint}[2][]{\url{#2}}

\bibitem[{\citenamefont{{Pitrou}}(2018)}]{PitrouPRE1}
\bibinfo{author}{\bibfnamefont{C.}~\bibnamefont{{Pitrou}}},
  \bibinfo{journal}{Phys. Rev. E} \textbf{\bibinfo{volume}{97}},
  \bibinfo{pages}{043115} (\bibinfo{year}{2018}), \eprint{1511.02331}.

\bibitem[{\citenamefont{Arne et~al.}(2009)\citenamefont{Arne, Marheineke,
  Meister, and Wegener}}]{Fraunhofer1}
\bibinfo{author}{\bibfnamefont{W.}~\bibnamefont{Arne}},
  \bibinfo{author}{\bibfnamefont{N.}~\bibnamefont{Marheineke}},
  \bibinfo{author}{\bibfnamefont{A.}~\bibnamefont{Meister}}, \bibnamefont{and}
  \bibinfo{author}{\bibfnamefont{R.}~\bibnamefont{Wegener}},
  \bibinfo{journal}{Berichte des Fraunhofer ITWM}
  \textbf{\bibinfo{volume}{167}} (\bibinfo{year}{2009}).

\bibitem[{\citenamefont{{Ribe}}(2004)}]{Ribe2004}
\bibinfo{author}{\bibfnamefont{N.~M.} \bibnamefont{{Ribe}}},
  \bibinfo{journal}{Royal Society of London Proceedings Series A}
  \textbf{\bibinfo{volume}{460}}, \bibinfo{pages}{3223} (\bibinfo{year}{2004}).

\bibitem[{\citenamefont{{Ribe} et~al.}(2006)\citenamefont{{Ribe}, {Habibi}, and
  {Bonn}}}]{Ribe2006}
\bibinfo{author}{\bibfnamefont{N.~M.} \bibnamefont{{Ribe}}},
  \bibinfo{author}{\bibfnamefont{M.}~\bibnamefont{{Habibi}}}, \bibnamefont{and}
  \bibinfo{author}{\bibfnamefont{D.}~\bibnamefont{{Bonn}}},
  \bibinfo{journal}{Physics of Fluids} \textbf{\bibinfo{volume}{18}},
  \bibinfo{pages}{084102} (\bibinfo{year}{2006}).

\bibitem[{\citenamefont{Arne et~al.}(2015)\citenamefont{Arne, Marheineke,
  Meister, and Wegener}}]{Fraunhofer2}
\bibinfo{author}{\bibfnamefont{W.}~\bibnamefont{Arne}},
  \bibinfo{author}{\bibfnamefont{N.}~\bibnamefont{Marheineke}},
  \bibinfo{author}{\bibfnamefont{A.}~\bibnamefont{Meister}}, \bibnamefont{and}
  \bibinfo{author}{\bibfnamefont{R.}~\bibnamefont{Wegener}},
  \bibinfo{journal}{J. of Comput. Physics} \textbf{\bibinfo{volume}{294}},
  \bibinfo{pages}{20} (\bibinfo{year}{2015}).

\bibitem[{\citenamefont{Marheineke and Wegener}(2007)}]{Fraunhofer3}
\bibinfo{author}{\bibfnamefont{N.}~\bibnamefont{Marheineke}} \bibnamefont{and}
  \bibinfo{author}{\bibfnamefont{R.}~\bibnamefont{Wegener}},
  \bibinfo{journal}{Berichte des Fraunhofer ITWM}
  \textbf{\bibinfo{volume}{115}} (\bibinfo{year}{2007}).

\bibitem[{\citenamefont{{Marheineke} and {Wegener}}(2009)}]{Fraunhofer4}
\bibinfo{author}{\bibfnamefont{N.}~\bibnamefont{{Marheineke}}}
  \bibnamefont{and}
  \bibinfo{author}{\bibfnamefont{R.}~\bibnamefont{{Wegener}}},
  \bibinfo{journal}{J. of Fluid Mech.} \textbf{\bibinfo{volume}{622}},
  \bibinfo{pages}{345} (\bibinfo{year}{2009}).

\bibitem[{\citenamefont{Bechtel et~al.}(1988)\citenamefont{Bechtel, Lin, and
  Forest}}]{BLF}
\bibinfo{author}{\bibfnamefont{S.~E.} \bibnamefont{Bechtel}},
  \bibinfo{author}{\bibfnamefont{K.~J.} \bibnamefont{Lin}}, \bibnamefont{and}
  \bibinfo{author}{\bibfnamefont{M.~G.} \bibnamefont{Forest}},
  \bibinfo{journal}{J. of Non-Newtonian Fluid Mech.}
  \textbf{\bibinfo{volume}{27}}, \bibinfo{pages}{87} (\bibinfo{year}{1988}).

\bibitem[{\citenamefont{{Bowick} and {Yao}}(2011)}]{Yao10}
\bibinfo{author}{\bibfnamefont{M.}~\bibnamefont{{Bowick}}} \bibnamefont{and}
  \bibinfo{author}{\bibfnamefont{Z.}~\bibnamefont{{Yao}}},
  \bibinfo{journal}{Eur. Phys. J. E} \textbf{\bibinfo{volume}{34}},
  \bibinfo{pages}{32} (\bibinfo{year}{2011}), \eprint{1011.3437}.

\bibitem[{\citenamefont{{Darbois Texier} et~al.}(2013)\citenamefont{{Darbois
  Texier}, {Piroird}, {Qu{\'e}r{\'e}}, and {Clanet}}}]{Dar13}
\bibinfo{author}{\bibfnamefont{B.}~\bibnamefont{{Darbois Texier}}},
  \bibinfo{author}{\bibfnamefont{K.}~\bibnamefont{{Piroird}}},
  \bibinfo{author}{\bibfnamefont{D.}~\bibnamefont{{Qu{\'e}r{\'e}}}},
  \bibnamefont{and} \bibinfo{author}{\bibfnamefont{C.}~\bibnamefont{{Clanet}}},
  \bibinfo{journal}{J. of Fluid Mech.} \textbf{\bibinfo{volume}{717}},
  \bibinfo{eid}{R3} (\bibinfo{year}{2013}).

\bibitem[{\citenamefont{Chandrasekhar}(1959)}]{Cha59}
\bibinfo{author}{\bibfnamefont{S.}~\bibnamefont{Chandrasekhar}},
  \bibinfo{journal}{Proc. London Math. Soc.} \textbf{\bibinfo{volume}{9}},
  \bibinfo{pages}{141} (\bibinfo{year}{1959}).

\bibitem[{\citenamefont{Becker et~al.}(1994)\citenamefont{Becker, Hiller, and
  Kowalewski}}]{Bec94}
\bibinfo{author}{\bibfnamefont{E.}~\bibnamefont{Becker}},
  \bibinfo{author}{\bibfnamefont{W.~J.} \bibnamefont{Hiller}},
  \bibnamefont{and} \bibinfo{author}{\bibfnamefont{T.~A.}
  \bibnamefont{Kowalewski}}, \bibinfo{journal}{J. of Fluid Mech.}
  \textbf{\bibinfo{volume}{258}}, \bibinfo{pages}{191} (\bibinfo{year}{1994}).

\bibitem[{\citenamefont{{Rayleigh}}(1878)}]{Rayleigh1878}
\bibinfo{author}{\bibfnamefont{J.~W.~S.} \bibnamefont{{Rayleigh}}},
  \bibinfo{journal}{Proc. R. Soc. London} \textbf{\bibinfo{volume}{10}},
  \bibinfo{pages}{4} (\bibinfo{year}{1878}).

\bibitem[{\citenamefont{Plateau}(1873)}]{Plateau1873}
\bibinfo{author}{\bibfnamefont{J.}~\bibnamefont{Plateau}},
  \emph{\bibinfo{title}{{Statique exp\'erimentale et th\'eorique des liquides
  soumis aux seules forces mol\'eculaires}}} (\bibinfo{publisher}{Paris,
  Gauthier-Villars}, \bibinfo{year}{1873}).

\bibitem[{\citenamefont{Rayleigh}(1892)}]{Rayleigh1892}
\bibinfo{author}{\bibfnamefont{L.}~\bibnamefont{Rayleigh}},
  \bibinfo{journal}{Philos. Mag.} \textbf{\bibinfo{volume}{34}},
  \bibinfo{pages}{145} (\bibinfo{year}{1892}).

\bibitem[{\citenamefont{Weber}(1931)}]{Weber1931}
\bibinfo{author}{\bibfnamefont{C.}~\bibnamefont{Weber}}, \bibinfo{journal}{Z.
  Angew. Math. Mech.} \textbf{\bibinfo{volume}{11}}, \bibinfo{pages}{136}
  (\bibinfo{year}{1931}).

\bibitem[{\citenamefont{{Eggers}}(1997)}]{EggersRMP}
\bibinfo{author}{\bibfnamefont{J.}~\bibnamefont{{Eggers}}},
  \bibinfo{journal}{Reviews of Modern Physics} \textbf{\bibinfo{volume}{69}},
  \bibinfo{pages}{865} (\bibinfo{year}{1997}).

\bibitem[{\citenamefont{{Eggers} and {Villermaux}}(2008)}]{Eggers2008}
\bibinfo{author}{\bibfnamefont{J.}~\bibnamefont{{Eggers}}} \bibnamefont{and}
  \bibinfo{author}{\bibfnamefont{E.}~\bibnamefont{{Villermaux}}},
  \bibinfo{journal}{Reports on Progress in Physics}
  \textbf{\bibinfo{volume}{71}}, \bibinfo{eid}{036601} (\bibinfo{year}{2008}).

\bibitem[{\citenamefont{{Driessen} et~al.}(2013)\citenamefont{{Driessen},
  {Jeurissen}, {Wijshoff}, {Toschi}, and {Lohse}}}]{Dri13}
\bibinfo{author}{\bibfnamefont{T.}~\bibnamefont{{Driessen}}},
  \bibinfo{author}{\bibfnamefont{R.}~\bibnamefont{{Jeurissen}}},
  \bibinfo{author}{\bibfnamefont{H.}~\bibnamefont{{Wijshoff}}},
  \bibinfo{author}{\bibfnamefont{F.}~\bibnamefont{{Toschi}}}, \bibnamefont{and}
  \bibinfo{author}{\bibfnamefont{D.}~\bibnamefont{{Lohse}}},
  \bibinfo{journal}{J. Phys. Fluids} \textbf{\bibinfo{volume}{25}},
  \bibinfo{pages}{062109} (\bibinfo{year}{2013}), \eprint{1307.3139}.

\bibitem[{\citenamefont{Garc\'ia and Castellanos}(1994)}]{GarciaCastellanos}
\bibinfo{author}{\bibfnamefont{F.~J.} \bibnamefont{Garc\'ia}} \bibnamefont{and}
  \bibinfo{author}{\bibfnamefont{A.}~\bibnamefont{Castellanos}},
  \bibinfo{journal}{Physics of Fluids} \textbf{\bibinfo{volume}{6}},
  \bibinfo{pages}{2676} (\bibinfo{year}{1994}).

\bibitem[{\citenamefont{{Pairam} and {Fern{\'a}ndez-Nieves}}(2009)}]{Pai09}
\bibinfo{author}{\bibfnamefont{E.}~\bibnamefont{{Pairam}}} \bibnamefont{and}
  \bibinfo{author}{\bibfnamefont{A.}~\bibnamefont{{Fern{\'a}ndez-Nieves}}},
  \bibinfo{journal}{Phys. Rev. Lett.} \textbf{\bibinfo{volume}{102}},
  \bibinfo{eid}{234501} (\bibinfo{year}{2009}).

\bibitem[{\citenamefont{{McGraw} et~al.}(2010)\citenamefont{{McGraw}, {Li},
  {Tran}, {Shi}, and {Dalnoki-Veress}}}]{McGraw15}
\bibinfo{author}{\bibfnamefont{J.~D.} \bibnamefont{{McGraw}}},
  \bibinfo{author}{\bibfnamefont{J.}~\bibnamefont{{Li}}},
  \bibinfo{author}{\bibfnamefont{D.~L.} \bibnamefont{{Tran}}},
  \bibinfo{author}{\bibfnamefont{A.-C.} \bibnamefont{{Shi}}}, \bibnamefont{and}
  \bibinfo{author}{\bibfnamefont{K.}~\bibnamefont{{Dalnoki-Veress}}},
  \bibinfo{journal}{Soft Matter} \textbf{\bibinfo{volume}{6}},
  \bibinfo{pages}{1258} (\bibinfo{year}{2010}).

\bibitem[{\citenamefont{Mehrabian and J.~Feng}(2013)}]{Meh13}
\bibinfo{author}{\bibfnamefont{H.}~\bibnamefont{Mehrabian}} \bibnamefont{and}
  \bibinfo{author}{\bibfnamefont{J.}~\bibnamefont{J.~Feng}},
  \bibinfo{journal}{J. of Fluid Mech.} \textbf{\bibinfo{volume}{717}},
  \bibinfo{pages}{281} (\bibinfo{year}{2013}).

\end{thebibliography}

\appendix

\section{Linearized viscous torus perturbations}\label{AppLinearEqs}

Dynamical equations for the variables (\ref{PerturbedVariables}) are
obtained from Eqs.~(7.18a), (7.18b), (7.20), and (7.42) and their ${\cal
  O}(\epsilon_\ra^2)$ corrections [Eqs.~(7.41), (G3), and (G4)] of
Ref.~\cite{PitrouPRE1}. After linearizing them, one uses the linearized constraints
(\ref{EqsConstraint2}) and (\ref{EqsConstraint3}) to get
\begin{widetext}
\bea
\partial_t \delta \ScV&=&-\Sv_0\partial_s \delta \ScV+U^r_0 \partial_s
\delta U^r-U^r_0 \kappa_0 \delta \ScV +3\visc \partial_s^2 \delta
\ScV+3 \visc \kappa_0 \partial_s \delta U^r+3 \visc U^r_0 \partial_s
\delta \kappa+6\visc U^r_0 \kappa_0 \frac{\partial_s \delta
  R}{R_0}+\frac{\ST \partial_s
\delta R}{R_0^2}\\
&&+\ST R_0\left(\tfrac{3}{4}\kappa_0^2\frac{\partial_s \delta
  R}{R_0}+\tfrac{1}{4}\kappa_0 \partial_s \delta
  \kappa+\frac{\partial_s^3 \delta R}{R_0}+ \frac{\partial_s^3\delta R}{R_0}\right)\nonumber\\
&&+R_0^2\Big( \tfrac{9}{4}U^r_0 \kappa_0^3 \delta \ScV 
+\tfrac{3}{4}(U^r_0)^2 \kappa_0^2 \frac{\partial_s \delta
  R}{R_0}-\tfrac{9}{2}\visc U^r_0 \kappa_0^3 \frac{\partial_s \delta
  R}{R_0}-\tfrac{15}{8}U^r_0 \kappa_0^2 \partial_s \delta
U^r -\tfrac{63}{16}\visc \kappa_0^3 \partial_s \delta
U^r+\tfrac{3}{8}(U^r_0)^2 \kappa_0 \partial_s \delta \kappa\nonumber\\
&&
\qquad -\tfrac{75}{16} \visc U^r_0 \kappa_0^2 \partial_s \delta
\kappa+\tfrac{3}{8} U^r_0\kappa_0\partial_s^2 \delta \ScV
-\tfrac{57}{16}\visc \kappa_0^2 \partial_s^2 \delta\ScV +3 \visc U^r_0 \kappa_0
\frac{\partial_s^3 \delta R}{R_0}+\tfrac{3}{8}\visc U^r_0 \partial_s^3
\delta \kappa + \tfrac{3}{8}\visc \partial_s^4 \delta
\ScV
\Big)\nonumber
\eea
\bea
\partial_t \delta U^r &=&-\Sv_0 \partial_s \delta U^r -\ST \frac{\delta \kappa}{R_0}+\ST \kappa_0
\frac{\delta R}{R_0^2}-6 \visc U^r_0 \kappa_0 \delta \kappa - 3 \visc
\kappa_0^2 \delta U^r -3 \visc \kappa_0 \partial_s \delta \ScV
\nonumber\\
&&+\ST R_0 \left(-\tfrac{21}{8}\kappa_0^2 \delta
\kappa-\tfrac{7}{8}\kappa_0^3 \frac{\delta
  R}{R_0}-\tfrac{13}{8}\kappa_0\frac{\partial_s^2 \delta
  R}{R_0}-\tfrac{3}{4}\partial_s^2 \delta \kappa\right)\nonumber\\
&&+R_0^2\Big(-\tfrac{63}{16} (U^r_0)^2 \kappa_0^2 \delta \kappa
-\tfrac{21}{8}(U^r_0)^2\kappa_0^3 \frac{\delta R}{R_0}
-\tfrac{21}{8}U^r_0 \kappa_0^3 \delta U^r -\tfrac{33}{4}\visc U^r_0
\kappa_0^3 \delta \kappa -\tfrac{33}{8} \visc U^r_0 \kappa_0^4
\frac{\delta R}{R_0}-\tfrac{33}{16}\visc \delta U^r
\kappa_0^4\nonumber\\
&&\qquad -\tfrac{3}{8}U^r_0 \kappa_0^2 \partial_s \delta \ScV
-\tfrac{27}{16}\visc \kappa_0^3 \partial_s \delta \ScV-\tfrac{39}{4}\visc U^r_0 \kappa_0^2 \frac{\partial_s^2 \delta
  R}{R_0}-\tfrac{9}{4} U^r_0\kappa_0 \partial_s^2 \delta U^r
-\tfrac{9}{2}\visc \kappa_0^2 \partial_s^2 \delta
U^r\nonumber\\
&&\qquad-\tfrac{27}{4}\visc U^r_0 \kappa_0 \partial_s^2 \delta \kappa
-\tfrac{27}{8}\visc \kappa_0 \partial_s^3 \delta \ScV
-\tfrac{3}{4}\visc \partial_s^4 \delta U^r
\Big)
\eea
\bea
\partial_t \delta \hrbp &=& -\Sv_0 \partial_s \delta \hrbp
+\hrbp_0 \partial_s \delta \ScV
+\frac{\ST}{\visc \ra_0}\left(\delta \hrbp +\frac{\hrbp_0 \delta R}{R_0}- \tfrac{1}{3} \kappa_0 \delta \kappa +\tfrac{1}{6} \frac{\kappa_0^2\delta R}{\ra_0}\right) 
+\hrbp_0
U^r_0\delta \kappa + U^r_0 \kappa_0 \delta \hrbp +\kappa_0 \hrbp_0
\delta U^r \nonumber\\
&&
-\tfrac{21}{16}U^r_0 \kappa_0^2 \delta \kappa -\tfrac{7}{16}\kappa_0^3
\delta U^r -\tfrac{5}{16}\kappa_0^2 \partial_s \delta \ScV
-\tfrac{1}{8}\kappa_0 \partial_s^2 \delta U^r
\eea
\bea
\frac{\partial_t \delta R}{R_0}&=&-\Sv_0 \frac{\partial_s \delta
  R}{R_0}-\tfrac{1}{2}U^r_0 \delta \kappa-\tfrac{1}{2}\kappa_0 \delta
U^r-\tfrac{1}{2}  \partial_s \delta \ScV +R_0^2 \left(-\tfrac{3}{8}U^r_0 \kappa_0
  \frac{\partial_s^2 \delta R}{R_0}-\tfrac{1}{16}\kappa_0 \partial_s^2 \delta
  U^r -\tfrac{1}{16}U^r_0 \partial_s^2 \delta \kappa  -\tfrac{1}{16} \partial_s^3 \delta \ScV\right)
\eea
\end{widetext}

\end{document}